\documentclass[runningheads]{llncs}
\usepackage[T1]{fontenc}
\usepackage{graphicx,verbatim}
\usepackage{hyperref}
\usepackage{color}

\usepackage{booktabs}
\usepackage{pifont}
\usepackage[normalem]{ulem}
\useunder{\uline}{\ul}{}
\usepackage{multirow}
\usepackage{amsmath}
\usepackage{amssymb}

\begin{document}
\title{Mono2D: A Trainable Monogenic Layer for
Robust Knee Cartilage Segmentation on Out-of-Distribution 2D Ultrasound Data
}
%
\author{Alvin Kimbowa\inst{1}
\and Arjun Parmar\inst{2}
\and Maziar Badii\inst{1}
\and David Liu\inst{1}
\and Matthew Harkey\inst{2}
\and Ilker Hacihaliloglu\inst{1}
}
\authorrunning{A. Kimbowa et al.}

%
\institute{The University of British Columbia, Vancouver, Canada
\and Michigan State University, Michigan, USA\\
\email{alvinbk@student.ubc.ca}
}



\maketitle              
\begin{abstract}
Automated knee cartilage segmentation using point-of-care ultrasound devices and deep-learning networks has the potential to enhance the management of knee osteoarthritis. 
However, segmentation algorithms often struggle with domain shifts caused by variations in ultrasound devices and acquisition parameters, limiting their generalizability.
In this paper, we propose Mono2D, a monogenic layer that extracts multi-scale, contrast- and intensity-invariant local phase features using trainable bandpass quadrature filters. 
This layer mitigates domain shifts, improving generalization to out-of-distribution domains.
Mono2D is integrated before the first layer of a segmentation network, and its parameters jointly trained alongside the network’s parameters.
We evaluated Mono2D on a multi-domain 2D ultrasound knee cartilage dataset for single-source domain generalization (SSDG).
Our results demonstrate that Mono2D outperforms other SSDG methods in terms of Dice score and mean average surface distance.
To further assess its generalizability, we evaluate Mono2D on a multi-site prostate MRI dataset, where it continues to outperform other SSDG methods, highlighting its potential to improve domain generalization in medical imaging.
Nevertheless, further evaluation on diverse datasets is still necessary to assess its clinical utility.
\textbf{The code will be made available after the review process.}



\end{abstract}
\section{Introduction}
Quantitative assessment and monitoring of knee cartilage health is crucial to improving our understanding and management of knee osteoarthritis~\cite{desai_knee-cartilage_2019}.
With the recent advancements in ultrasound technology, point-of-care ultrasound has gained prominence as a viable alternative imaging modality for the quantitative assessment and monitoring of cartilage health, particularly in individuals at high risk of developing osteoarthritis, within decentralized healthcare settings~\cite{nakashima_point--care_2022}.
However, this assessment relies on accurate cartilage segmentation which is currently mostly manual requiring a lot of expertise, time, and exhibits inter- and intra-rater variability.
Deep learning-based knee cartilage segmentation has shown promise for automating cartilage segmentation and thickness measurement~\cite{desai_knee-cartilage_2019,harkey_validating_2022}, however, these methods are typically computationally expensive and require access to high-performance computing infrastructure.
Recent efforts have focused on developing lightweight algorithms that can operate efficiently on-device, making them suitable for resource-limited environments where cloud-based high-performance computing is inaccessible~\cite{valanarasu_unext_2022}.
However, such algorithms often struggle to generalize across varying ultrasound image qualities obtained with different devices, acquisition settings, and operator variability, limiting their clinical utility.

The domain shift challenge can be solved by training with datasets from multiple domains including the desired target domain; however, such data is non-existent, particularly knee cartilage, and collecting it is expensive and may be infeasible in clinical scenarios.
Another approach is to perform data augmentation on the source domain to simulate potential domain shifts, and various single-source domain generalization (SSDG) approaches have been proposed.
For instance, Zhang et al. proposed using deep stacked transformations that alter image quality, appearance, and spatial configuration~\cite{zhang_generalizing_2020}.
Su et al. proposed a saliency-balancing and location-scale augmentation approach that uses gradient information to enrich the informativeness of the augmented images~\cite{su_rethinking_2023}.
Jiang and Gu use edge detectors, such as Sobel and Canny detectors, to extract edge information and pass this to the segmentation model~\cite{junjiang_train_2024}.
Ouyang et al. proposed using randomly-weighted shallow networks and simulated domain-specific perturbations to train a domain-invariant model that ignores these augmentations~\cite{ouyang_causality-inspired_2023}.
Arslan et al. proposed using a frequency-based Lipschitz regularization loss to limit the sensitivity of a model's latent features to high-frequency components~\cite{arslan_single-source_2024}.
Hu et al. took the contrastive approach where they use projection modules to disentangle the model's shallow features into style-augmented and structure representations~\cite{hu_devil_2023}.
Generally, these approaches can yield unrealistic knee ultrasound images whose effect on the model learning remains unclear, require expertise and knowledge of the possible domain shifts, maybe be unrealistic for knee cartilage ultrasound, or are challenging to incorporate into existing model training pipelines.

Local phase information has been shown to enhance ultrasound images to aid bone and knee cartilage segmentation~\cite{harkey_validating_2022}.
However, extracting local phase information typically relies on manual tuning of bandpass quadrature filters, such as the Log-Gabor filter which requires significant expertise~\cite{belaid2024,bridge_introduction_2017,hacihaliloglu_automatic_2011}.
Hacihaliloglu et al. proposed an automated approach for selecting hyperparameters of the monogenic layer at test time~\cite{hacihaliloglu_automatic_2011}, however, this approach requires multiple target domain images followed by an iterative approach that may not be feasible for real-time and or on-device segmentation.
Moya-Sánchez et al. proposed the first trainable monogenic layer for domain generalization~\cite{moya-sanchez_trainable_2021}.
However, their approach only extracts single-scale local-phase and orientation information, and the layer parameters are unconstrained making it numerically unstable during training.
Furthermore, the layer was evaluated only on synthetically degraded natural images.

\begin{figure}[t]
    \centering    
    \includegraphics[width=\linewidth]{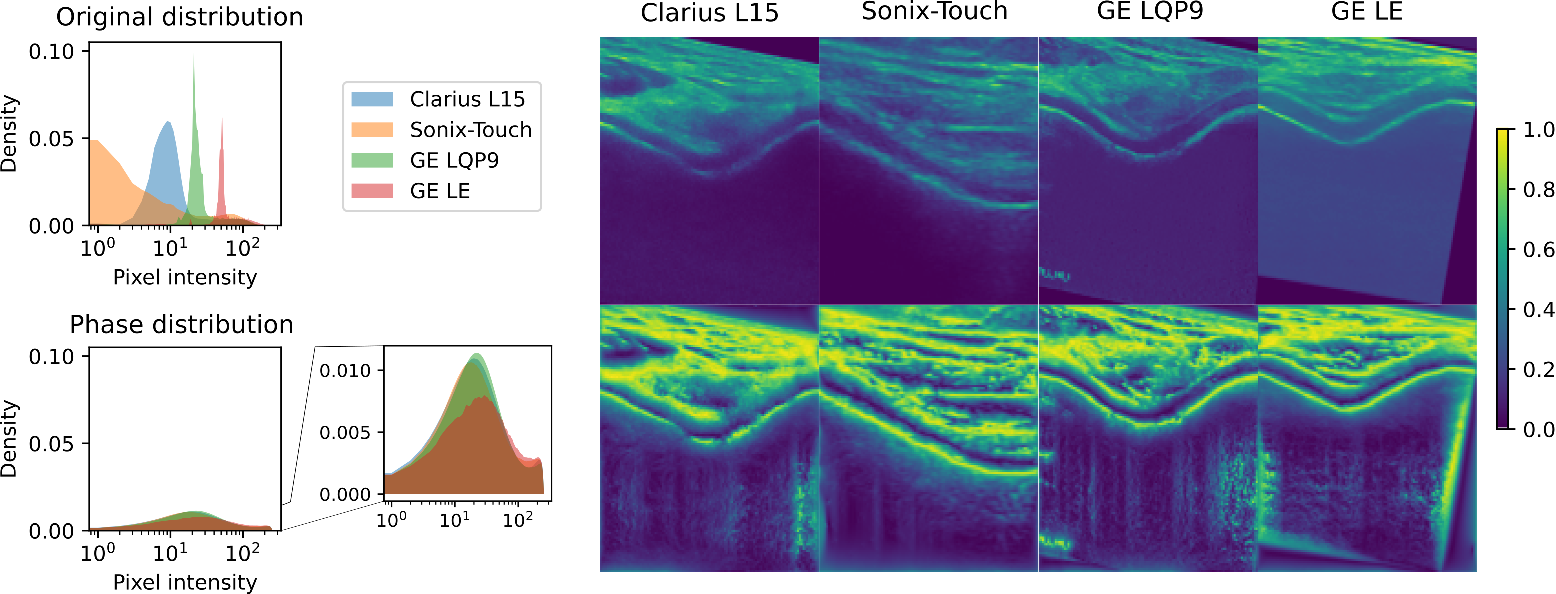}
    \caption{
    \textbf{Left:} The domain shift in the original datasets (top histogram) can be mitigated using local phase information (bottom histogram).
    \textbf{Right:}
    The relatively unclear cartilage structure in the original images (top row) is uniformly enhanced in the local phase images across all datasets (bottom row).
    }
    \label{fig:datasets}
\end{figure}

In this paper, we propose Mono2D, a trainable monogenic layer, that automatically learns the bandwidths and scales of the Log-Gabor filters for optimizing local phase features.
By automatically learning these features, we enhance generalization to domain changes, as local phase features are invariant to image intensity and contrast variations, and enhance structural information (cf. Fig.~\ref{fig:datasets}).
Evaluation on a multi-domain 2D ultrasound knee cartilage dataset indicates that Mono2D outperforms other SSDG methods.
Further evaluation on a multi-site prostate MRI dataset demonstrates Mono2D's potential to generalize to other medical imaging modalities.

\section{Methodology}\label{methodology}



\subsection{Mono2D layer}

\subsubsection{Background} A monogenic signal is a mathematical representation that decomposes an image into structural information---local phase and orientation---and energetic information---local energy~\cite{felsberg_monogenic_2001}.
The local phase contains the majority of the information required for image analysis~\cite{felsberg_monogenic_2001}.
Additionally, other phase-based features such as phase asymmetry, which enhances edges of structures, can be computed from the local phase information~\cite{felsberg_monogenic_2001,kovesi_symmetry_nodate}.
Formally, the monogenic signal of an image $I(x,y)$ is defined as a triplet $\mathbf{M}(I(x,y)) = (I_f, R_{1}, R_{2})$, where $I_f$ is a bandpass filtered version of $I(x,y)$, and $R_1 = R_x(I_f)$ and $R_2 = R_y(I_f)$ are the even and odd Riesz transforms of $I_f$, respectively.
The local phase, $I_\theta$, and phase asymmetry of the signal at a given point in the image is defined by
\begin{align}
    I_\theta = arctan\left( \frac{I_f}{R_1^2 + R_2^2}\right),
    I_{asym} = \frac{\lfloor |R_1,R_2| - |I_f| \rfloor}{|I_f,R_1,R_2|}
    \label{eqn:local_phase}
\end{align}

\subsubsection{Layer definition}
Computing the monogenic signal is easier in the frequency domain~\cite{felsberg_monogenic_2001}; therefore, the Mono2D layer starts by transforming the input image to the frequency domain using a fast Fourier Transform (FFT).
A low-pass filter (LPF) is then applied to the FFT response to eliminate extra high frequencies that can distort certain normalization processes during phase information computation~\cite{belaid2024}.
We use a Butterworth filter with a normalized cutoff frequency of 0.5 and order 10 as in~\cite{belaid2024}.
Since the monogenic signal is not selective in frequency, a bandpass filter is applied to extract features at a given scale.
A popular choice is the log-Gabor filter (LGF) parameterized by the center frequency $f_0$, which specifies the scale, and the relative bandwidth $\sigma_r$ that controls the selectivity of the filter~\cite{belaid2024,bridge_introduction_2017}.
The LGF response $G$ is given by
\begin{align}
    G(f) &= \exp \left( - \frac{ \left( \log\left(\frac{f}{f_0}\right) \right)^2 }{ 2 \left( \log\left(\sigma_r\right) \right)^2 } \right),
    \label{eqn:log-gabor}
\end{align}
where $f$ is the spatial frequency.
Note that $f_0$ and $\sigma_r$ are the only trainable parameters of the Mono2D layer. 
For multi-scale feature extraction, $n$ bandpass filters are computed in parallel and their responses are summed~\cite{bridge_introduction_2017}.
The $x$ and $y$ direction Riesz kernels, $R_x$, and $R_y$, are then applied to the LGF response to compute the odd parts from the original signal.
The responses of the filters can be combined into a single complex number given by
\begin{align}
    iR_x - R_y = \frac{if_x - f_y}{\sqrt{f_x^2 + f_y^2}}.
\end{align}
at each point in the image, where $f_x$ and $f_y$ are the $x$ and $y$ frequencies, respectively.
Taking the inverse Fourier Transform (IFFT) of the responses from the LGF and Riesz kernels yields the monogenic signal consisting of three components, $I_f$, $R_1$, and $R_2$.
The local phase and phase asymmetry are then computed following equation~\ref{eqn:local_phase}.

\begin{figure}[t]
    \centering
    \includegraphics[width=0.9\linewidth]{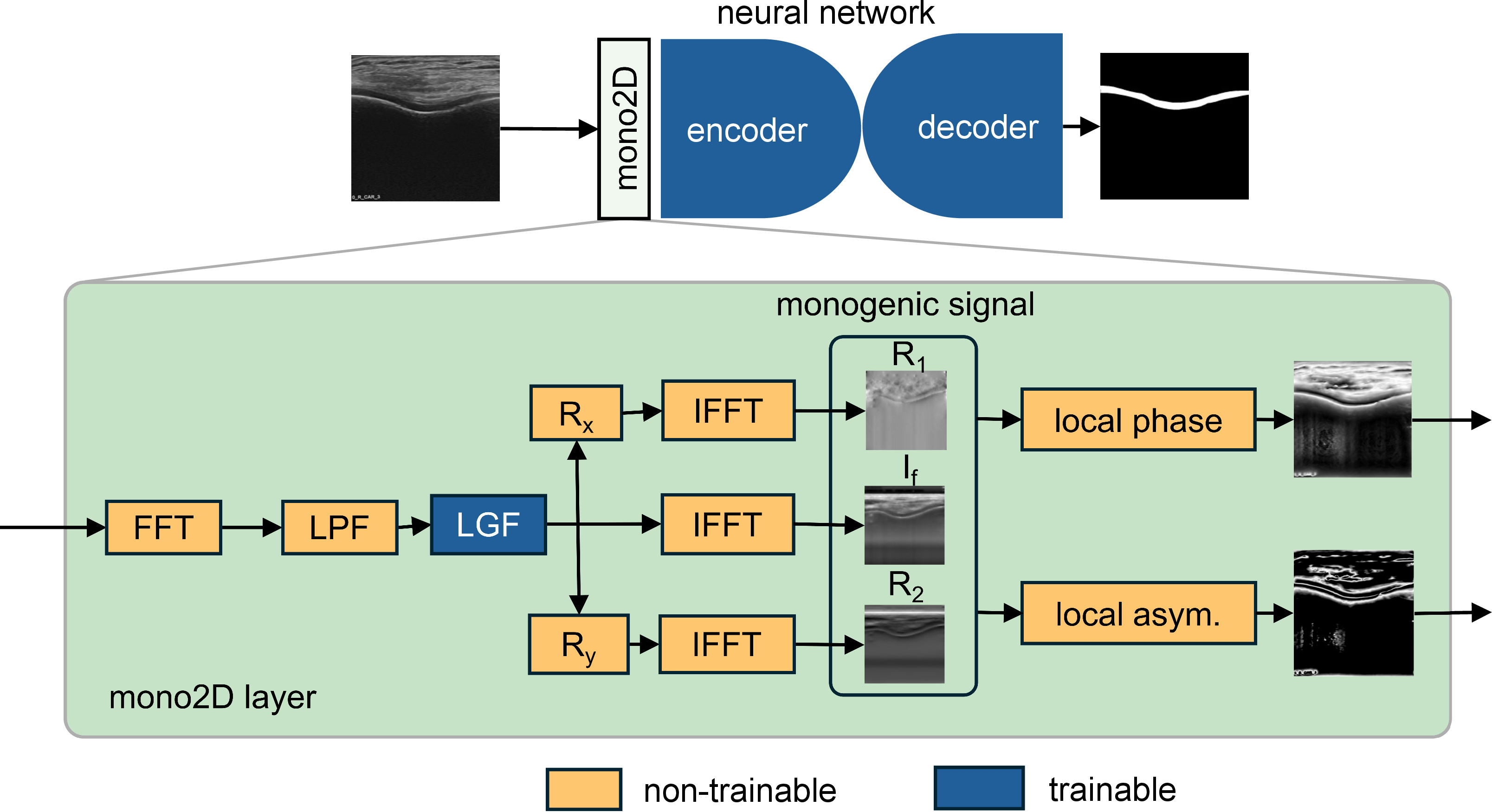}
    \caption{\textbf{Approach overview:} The Mono2D layer is placed before a neural network. The layer consists of a fast Fourier Transform (FFT), followed by a low-pass filter (LPF), log-Gabor filter (LGF), Riesz kernels $R_x$ and $R_y$, and inverse FFT (IFFT).
    }
    \label{fig:approach}
\end{figure}

In summary, for an input image $I$, the Mono2D layer operations can be summarized as;
\begin{align}
    \mathbf{I}_{filtered} &= \mathcal{F}\{I\} * LPF * LGF,\\
    I_f &= \mathcal{F}^{-1}\{\mathbf{I}_{filtered}\},\\
    R_1 + iR_2 &= \mathcal{F}^{-1}\{\mathbf{I}_filtered * (iR_x - R_y\},\\
    I_{out} &= I_\theta(I_f, R_1, R_2) + 
\end{align}
where $\mathbf{I}_{filtered}$ is the bandpass filtered image in the frequency domain.
For all experiments in this paper, we use the min-max scaler to rescale the Mono2D output to $[0,1]$ as most machine learning algorithms expect inputs in this range.
Fig.~\ref{fig:approach} shows the integration of Mono2D into a neural network.

\subsubsection{Layer training}
Directly optimizing the log-Gabor filter parameters, $f_0$ and $\sigma_r$, is numerically unstable if the optimizer drives the values of parameters outside their valid ranges; $f_0>=0$, and $0<\sigma_r<=1$.
We address this challenge by optimizing numerically stable unbounded parameters and deriving the actual parameters from them.
For $f_0$, we optimize an unbounded parameter $f_0^*$ that is related to $f_0$ by 
\begin{align}
    f_0 = f_{0_{min}} + sigmoid(f_0^*) \cdot (f_{0_{max}} - f_{0_{min}}),
\end{align}
where $f_{0_{min}}$ and $f_{0_{max}}$ are the possible minimum and maximum spatial frequencies in the image, respectively, $f_0^*\in \mathbb{R}$, $f_{0_{min}} < f_0 < f_{0_{max}}$.
According to Nyquist theorem,
$f_{0_{min}}=1/max(H, W)$ cycles/pixel and $f_{0_{max}} = 0.5$ cycles/pixel where $H$ and $W$ are the image height and width respectively.
We initialize $f_0^*$ with a standard normal distribution to allow sampling different initial values.
Similarly, for $\sigma_r$, we optimize an unbounded parameter $\sigma_r^*$ related to $\sigma_r$ by
\begin{align}
    \sigma_r = sigmoid(\sigma_r^*),
\end{align}
where $\sigma_r^* \in \mathbb{R}$, and $0 < \sigma_r < 1$.
We initialize $\sigma_r^*$ with a normal distribution with zero mean and standard deviation 0.05 to keep the initial value at about 0.5 as in~\cite{belaid2024}.

\subsection{Experimental Setup}
\subsubsection{Datasets}
We used a retrospective local 2D ultrasound knee cartilage dataset, collected after consent from healthy subjects and those who underwent anterior cruciate ligament reconstruction (ACLR).
The dataset includes images from four ultrasound scanners: the GE LOGIQ P9 R3 ultrasound system with the L3-12-RS wideband linear array probe (GE Healthcare), the LOGIQe US system (General Electric Co., Fairfield, CT) with a 12 MHz linear probe, and the Clarius HD3 L15 scanner. 
For brevity, we assign IDs to the datasets as D1 (1787 images), D2 (587 images), and D3 (234 images), respectively.
To test the generalizability of Mono2D to other datasets, we also used the open-source multi-site prostate MRI segmentation dataset from six clinical sites: NCI-ISBI-2013~\cite{bloch_nci-isbi_2015}, I2CVB~\cite{lemaitre_computer-aided_2015}, and PROMISE12~\cite{litjens_evaluation_2014}.
We adopted the same preprocessing as in~\cite{liu2021feddg} followed by resizing the 2D slices to 256x256 and rescaling to the range [0, 1].

\subsubsection{Experiments}
We used UNeXt~\cite{valanarasu_unext_2022}, a lightweight medical image segmentation network, as the segmentation backbone due to its low computational complexity.
For each domain, we split the data into a 90:10 train:validation split, training on a single domain and testing on other domains using the model that yielded the best performance on the in-domain validation split.
To establish the lower-bound SSDG performance, we trained the network without any SSDG method and obtained its out-of-distribution performance.
For the upper-bound SSDG performance, we trained the network on the other domains and computed the average performance.
We then integrated Mono2D into the network and empirically set the number of scales $n$ to 8, as it yielded the best performance and aligned with the number of scales used in practice~\cite{belaid2024}.
We also used standard data augmentation transforms: rotation ($\pm20$\textdegree), scaling ($0.4, 1.6$), and translation ($15$ pixels) with inspiration from~\cite{zhang_generalizing_2020}.
We compared the performance of Mono2D with six recent SSDG methods: SLAug~\cite{su_rethinking_2023}, EGSDG~\cite{junjiang_train_2024}, CSDG~\cite{ouyang_causality-inspired_2023}, CCSDG~\cite{hu_devil_2023}, BigAug~\cite{zhang_generalizing_2020}, and LRFS~\cite{arslan_single-source_2024}.
We use the Dice similarity score and mean average surface distance (MASD) as the evaluation metrics following recommendations by Maier-Hein et al.~\cite{reinke_understanding_2024}.


\subsubsection{Implementation details}
All methods were implemented using the PyTorch framework and integrated into the training pipeline of UNeXt based on their papers and open-source implementations for fair comparison.
We used Adam optimizer with a learning rate of $1\times10{-3}$ and momentum of 0.9.
We also use a cosine annealing learning rate scheduler with a minimum learning rate of $1\times10^{-5}$ and train for 500 epochs with inspiration from the original UNeXt training pipeline.
Training was performed for 500 epochs while saving the best-performing model on the validation set during training.
We trained on a single NVIDIA Tesla V100 12GB graphical processing unit (GPU) with a batch size of 8.


\section{Results and Discussion}\label{results}

\begin{figure}[h]
    \centering
    \includegraphics[width=0.79\linewidth]{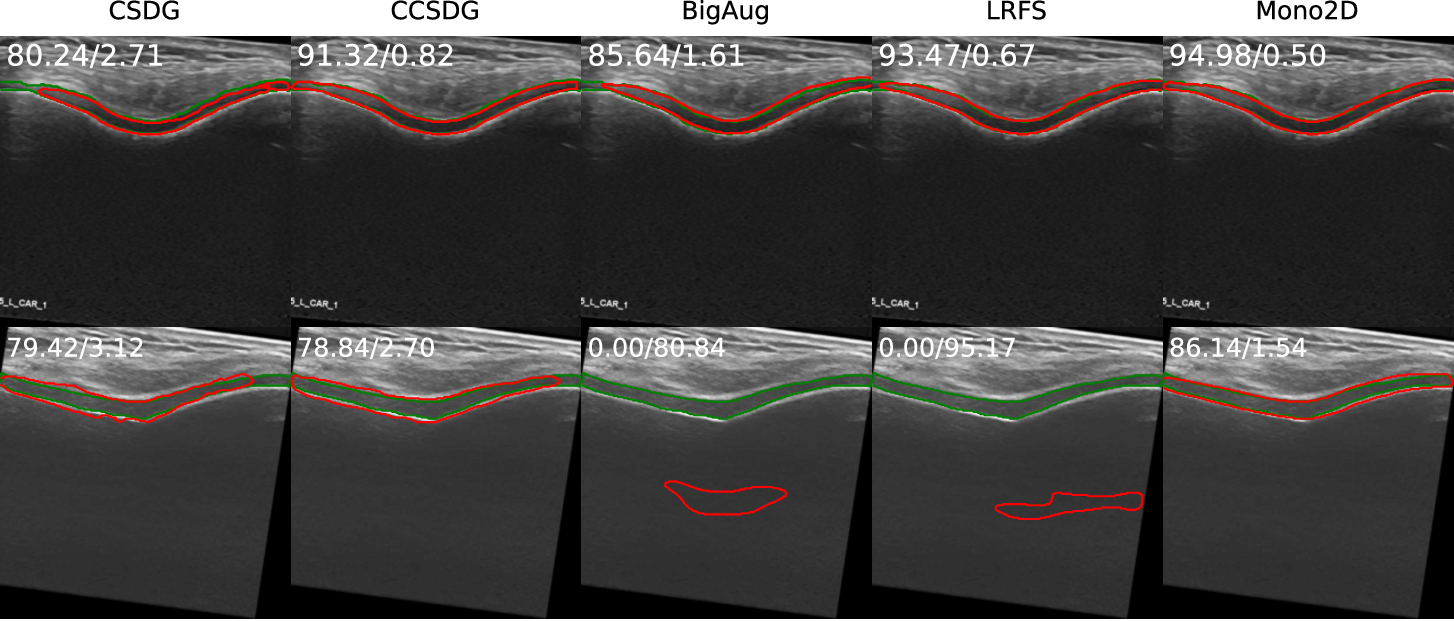}
    \caption{Sample qualitative results on unseen out-of-distribution images. The green contour indicates the expert label, and the red contour indicates the model prediction. 
    }
    \label{fig:qualitative_results}
\end{figure}

\begin{figure}[h]
    \centering
    \includegraphics[width=0.95\linewidth]{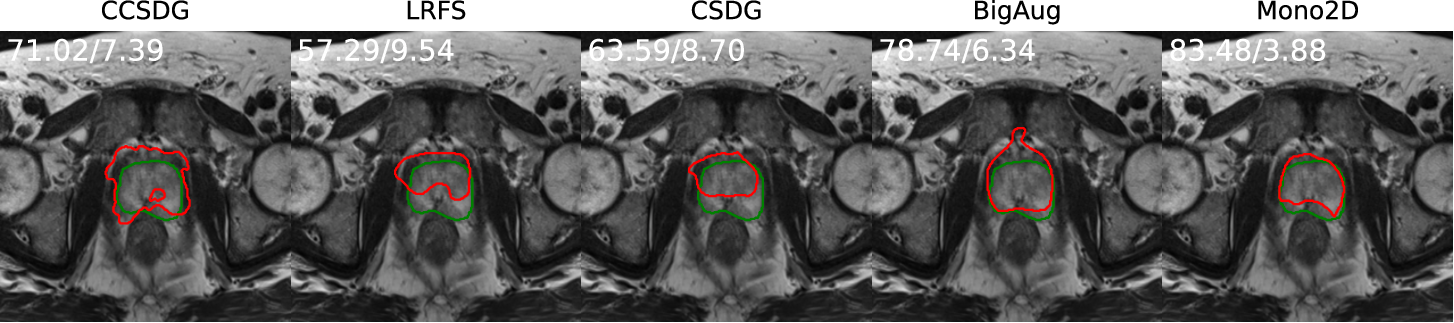}
    \caption{Sample qualitative results on unseen out-of-distribution images. The green contour indicates the expert label, and the red contour indicates the model prediction.}
    \label{fig:prostate}
\end{figure}

We tabulate the average percentage Dice and MASD for all methods in Table~\ref{tab:quantitative_results}.
Mono2D consistently outperforms other methods across all domains, achieving the highest average Dice score of 94.21\% and unit MASD of 1.28. 
As shown in Fig.~\ref{fig:qualitative_results}, Mono2D yields more complete segmentation of the cartilage.
Notably, LRFS and BigAug show unexpected failures, likely due to unrealistic augmentations that bias the model toward learning spurious features.
To validate the effectiveness of different Mono2D components, we performed an ablation study including freezing the layer, using only local phase, phase asymmetry, or both.
Results indicate that using training Mono2D yields better performance, and that both local phase and phase asymmetry outperforms using either feature alone.
Mono2D also exhibited similar performance on the multi-site prostate MRI dataset (Table~\ref{tab:prostate} and Fig.~\ref{fig:prostate}).

\begin{table}[t]
\centering
\caption{Performance comparison of Mono2D and existing methods in terms of Dice(\%)/MASD.
The best results are in bold and the second best are underlined. 
}
\label{tab:quantitative_results}
\begin{tabular}{@{}lcccccc@{}}
\toprule
\multicolumn{3}{l|}{\textbf{Model}} & \textbf{D1 $\rightarrow$ rest} & \textbf{D2 $\rightarrow$ rest} & \multicolumn{1}{c|}{\textbf{D3 $\rightarrow$ rest}} & \textbf{Avg} \\ \midrule
\multicolumn{3}{l|}{Lower bound} & 85.54/10.49 & 66.70/19.63 & \multicolumn{1}{c|}{77.40/39.51} & 76.55/23.21 \\ \midrule
\multicolumn{3}{l|}{SLAug~\cite{su_rethinking_2023}} & 92.36/1.76 & 91.99/1.69 & \multicolumn{1}{c|}{89.86/2.43} & 91.40/1.96 \\
\multicolumn{3}{l|}{EGSDG~\cite{junjiang_train_2024}} & 92.70/1.88 & 92.95/1.57 & \multicolumn{1}{c|}{90.78/2.20} & 92.15/1.88 \\
\multicolumn{3}{l|}{CSDG~\cite{ouyang_causality-inspired_2023}} & 92.94/1.70 & 93.58/1.38 & \multicolumn{1}{c|}{91.15/2.29} & 92.56/1.79 \\
\multicolumn{3}{l|}{CCSDG~\cite{hu_devil_2023}} & 91.45/2.41 & {\ul 94.01/1.31} & \multicolumn{1}{c|}{{\ul 93.08/1.54}} & 92.85/1.75 \\
\multicolumn{3}{l|}{BigAug~\cite{zhang_generalizing_2020}} & {\ul 94.03/1.38} & 93.86/1.35 & \multicolumn{1}{c|}{91.46/2.15} & 93.11/1.63 \\
\multicolumn{3}{l|}{LRFS~\cite{arslan_single-source_2024}} & 93.91/1.40 & {93.30/1.43} & \multicolumn{1}{c|}{92.79/1.60} & {\ul 93.33/1.48} \\
\multicolumn{3}{l|}{\textbf{Mono2D (Ours)}} & \textbf{94.27/1.32} & \textbf{94.70/1.13} & \multicolumn{1}{c|}{\textbf{93.64/1.39}} & \textbf{94.21/1.28} \\ \midrule
\multicolumn{3}{l|}{Upper bound} & 95.56/1.01 & 95.72/0.90 & \multicolumn{1}{c|}{95.10/1.04} & 95.46/0.98 \\ \midrule
\multicolumn{7}{c}{\textbf{Ablation study}} \\ \midrule
\multicolumn{1}{c}{\textbf{Trainable}} & \textbf{Phase} & \multicolumn{1}{c|}{\textbf{Asym.}} & \textbf{D1 $\rightarrow$ rest} & \textbf{D2 $\rightarrow$ rest} & \multicolumn{1}{c|}{\textbf{D3 $\rightarrow$ rest}} & \textbf{Avg} \\ \midrule
\multicolumn{1}{c}{} & \ding{51} & \multicolumn{1}{c|}{} & 94.31/1.32 & 94.47/1.19 & \multicolumn{1}{c|}{93.20/1.53} & 93.99/1.35 \\
\multicolumn{1}{c}{\ding{51}} & \ding{51} & \multicolumn{1}{c|}{} & \textbf{94.46/1.28} & 94.65/1.15 & \multicolumn{1}{c|}{93.39/1.51} & 94.16/1.31 \\
\multicolumn{1}{c}{\textbf{\ding{51}}} & \textbf{} & \multicolumn{1}{c|}{\textbf{\ding{51}}} & \textbf{94.38/1.29} & \textbf{94.69/1.14} & \multicolumn{1}{c|}{\textbf{93.30/1.53}} & \textbf{94.12/1.32} \\
\multicolumn{1}{c}{\ding{51}} & \ding{51} & \multicolumn{1}{c|}{\ding{51}} & 94.27/1.32 & \textbf{94.70/1.13} & \multicolumn{1}{c|}{\textbf{93.64/1.39}} & \textbf{94.21/1.28} \\ \bottomrule
\end{tabular}
\end{table}

\begin{table}[h]
\centering
\caption{Performance comparison of Mono2D and existing methods in terms of Dice(\%).
The best results are in bold and the second best are underlined. }
\label{tab:prostate}
\begin{tabular}{@{}l|llllll|l@{}}
\toprule
\textbf{Model} & \multicolumn{1}{c}{\textbf{A}} & \multicolumn{1}{c}{\textbf{B}} & \multicolumn{1}{c}{\textbf{C}} & \multicolumn{1}{c}{\textbf{D}} & \multicolumn{1}{c}{\textbf{E}} & \multicolumn{1}{c|}{\textbf{F}} & \multicolumn{1}{c}{\textbf{Avg}} \\ \midrule
Lower bound & 68.28 & 73.24 & 40.48 & 63.54 & 30.81 & 37.83 & 52.36 \\ \midrule
EGSDG~\cite{junjiang_train_2024} & 48.55 & 23.60 & 54.04 & 48.75 & 51.50 & 49.75 & 46.03 \\
SLAug~\cite{su_rethinking_2023} & \multicolumn{1}{c}{78.42} & \multicolumn{1}{c}{76.92} & \multicolumn{1}{c}{58.59} & \multicolumn{1}{c}{57.76} & \multicolumn{1}{c}{56.60} & \multicolumn{1}{c|}{55.12} & \multicolumn{1}{c}{63.90} \\
CCSDG~\cite{ouyang_causality-inspired_2023} & \multicolumn{1}{c}{75.57} & \multicolumn{1}{c}{79.59} & \multicolumn{1}{c}{35.84} & \multicolumn{1}{c}{\textbf{76.91}} & \multicolumn{1}{c}{64.93} & \multicolumn{1}{c|}{61.39} & \multicolumn{1}{c}{65.71} \\
LRFS~\cite{arslan_single-source_2024} & \multicolumn{1}{c}{79.65} & \multicolumn{1}{c}{68.24} & \multicolumn{1}{c}{58.03} & \multicolumn{1}{c}{72.74} & \multicolumn{1}{c}{{\ul 69.18}} & \multicolumn{1}{c|}{67.50} & \multicolumn{1}{c}{69.22} \\
CSDG~\cite{ouyang_causality-inspired_2023} & \multicolumn{1}{c}{77.44} & \multicolumn{1}{c}{82.51} & \multicolumn{1}{c}{63.94} & \multicolumn{1}{c}{69.62} & \multicolumn{1}{c}{62.44} & \multicolumn{1}{c|}{69.32} & \multicolumn{1}{c}{70.88} \\
BigAug~\cite{zhang_generalizing_2020} & {\ul 80.90} & {\ul 82.15} & {\ul 68.77} & {\ul 73.47} & 60.89 & 69.92 & 72.69 \\ 
\midrule
Mono2D (asym.) & 80.51 & 81.64 & 67.06 & 72.75 & 62.92 & {\ul 72.53} & 72.90 \\
Mono2D (phase) & \textbf{81.13} & \textbf{82.56} & 67.31 & 73.71 & \textbf{70.56} & 67.99 & {\ul 73.88} \\
\textbf{Mono2D (asym. + phase)} & \multicolumn{1}{c}{79.99} & \multicolumn{1}{c}{82.07} & \multicolumn{1}{c}{\textbf{71.08}} & \multicolumn{1}{c}{73.00} & \multicolumn{1}{c}{68.85} & \multicolumn{1}{c|}{\textbf{75.32}} & \multicolumn{1}{c}{\textbf{75.05}} \\ \midrule
Upper bound & 90.72 & 91.00 & 91.59 & 90.99 & 91.32 & 90.97 & 91.10 \\ \bottomrule
\end{tabular}
\end{table}

\section{Conclusion}\label{results}
In this paper, we introduce Mono2D, a trainable monogenic layer for single-source domain generalization in lightweight deep learning segmentation networks.
Mono2D outperformed other single-source domain generalization methods on a multi-domain 2D ultrasound knee cartilage dataset, with transferable performance on a multi-site prostate MRI dataset.
Unlike other SSDG methods, Mono2D can easily be integrated into any trainable network with ideally no changes to existing training pipelines and architectures.
To the best of our knowledge, this is the first work to investigate the use of a trainable monogenic layer for domain generalization in medical imaging.
Nevertheless, further evaluation on different medical imaging datasets, architectures, and tasks is still required.
Future work will also investigate optimizing other phase-based features, such as phase congruency, and integrating Mono2D into deeper layers of segmentation networks.
Future work will also evaluate other quadrature filters, such as alpha scale space derivative filters.

\bibliographystyle{splncs04}
\bibliography{mybibliography}

\clearpage

\end{document}